\documentclass[aps,prb,twocolumn,superscriptaddress]{revtex4}
\usepackage{graphicx}
\usepackage{bm}

\newcommand{ \etal }{\mbox{\sl et al. }}

\newcommand{\tbmno}{\mbox{TbMnO$_{3}$}}
\newcommand{\dymno}{\mbox{DyMnO$_{3}$}}

\newcommand{ \tndy}{\mbox{$T_N^{Dy}$}}
\newcommand{ \tntb }{\mbox{$T_N^{Tb}$}}

\newcommand{ \hca }{\mbox{$H^{a}_{C}$}}
\newcommand{ \hcb}{\mbox{$H^{b}_{C}$}}
\newcommand{ \mhca }{\mbox{$\mu_0 H^{a}_{C}$}}
\newcommand{ \mhcb}{\mbox{$\mu_0H^{b}_{C}$}}

\newcommand{ \deltam }{\mbox{$\delta_m$}}
\newcommand{ \deltal}{\mbox{$\delta_l$}}

\newcommand{ \deltamn }{\mbox{$\delta^{Mn}_m$}}

\newcommand{ \cdirection }{\mbox{c$^*$-direction}}

\newcommand{ \bdirection}{\mbox{b$^*$-direction}}
\newcommand{ \ppa }{\mbox{$P\|a$}}
\newcommand{ \ppc }{\mbox{$P\|c$}}
\newcommand{ \hpa}{\mbox{$H\|a$}}
\newcommand{ \hpb }{\mbox{$H\|b$}}
\newcommand{ \hpc}{\mbox{$H\|c$}}
\newcommand{ \mhpa}{\mbox{$\mu_0H\|a$}}
\newcommand{ \mhpb }{\mbox{$\mu_0H\|b$}}

\newcommand{ \tc}{\mbox{$T_C(Dy)$}}

\newcommand{ \Pol }{\boldmath${P}$\unboldmath}
\newcommand{ \hh }{\boldmath${H}$\unboldmath}
\newcommand{ \vecE }{\boldmath${e}$\unboldmath}

\begin{document}

\title{Absence of commensurate ordering at the polarization flop transition
in multiferroic DyMnO$_{3}$}
\author{J. Strempfer}
\affiliation{Hamburger Synchrotronstrahlungslabor HASYLAB at Deutsches
Elektronen-Synchrotron DESY, 22605 Hamburg, Germany}
\author{B. Bohnenbuck}
\affiliation{Max-Planck-Institut f\"ur Festk\"orperforschung, 70569 Stuttgart, Germany}
\author{M. Mostovoy}
\affiliation{Zernike Institute for Advanced Materials, University of Groningen, 9747 AG Groningen,
Netherlands}
\author{N. Aliouane}
\author{D.N. Argyriou}
\affiliation{Hahn-Meitner-Institut, 14109 Berlin, Germany}
\author{F. Schrettle}
\author{J. Hemberger}
\author{A. Krimmel}
\affiliation{Experimentalphysik V, Universit\"at Augsburg, 86159 Augsburg, Germany}
\author{M. v. Zimmermann}
\affiliation{Hamburger Synchrotronstrahlungslabor HASYLAB at Deutsches
Elektronen-Synchrotron DESY, 22605 Hamburg, Germany}
\date{\today}

\begin{abstract}
Ferroelectric spiral magnets \dymno\ and \tbmno\ show similar
behavior of electric polarization in applied magnetic fields.
Studies of the field dependence of lattice modulations on the
contrary show a completely different picture. Whereas in \tbmno\ the
polarization flop from \ppc\ to \ppa\ is accompanied by a sudden
change from incommensurate to commensurate wave vector modulation,
in \dymno\ the wave vector varies continuously through the flop
transition. This smooth behavior may be related to the giant
magnetocapacitive effect observed in \dymno.

\end{abstract}

\maketitle









The ability to control ferroelectric polarization (\Pol) with an
applied magnetic field (\hh) in manganite perovskites and related
materials has caused a resurgence of interest in magneto-electric
phenomena. One of the most striking effects is the five-fold increase
of the dielectric constant of \dymno\ in magnetic field, named giant
magnetocapacitance.\cite{Got04} The key to this unprecedented
sensitivity is the spiral magnetic ordering stabilized by competing exchange
interactions, which forces positive and
negative ions to shift in opposite directions and which can be
rather easily influenced by applied magnetic fields.

Unlike conventional ferroelectrics, in the $R$MnO$_{3}$ manganite
perovskites such as $R$=Tb and Dy the emergence of
ferroelectricity\cite {Kim03a,Kim05} arises from the peculiar
coupling of the lattice to a spiral ordering of
Mn-spins.\cite{Ken05,Kat05,Mos06} Spiral ordering is defined by the
wave vector \boldmath $\kappa$\unboldmath\ and the axis
\boldmath${e}$\unboldmath\ around which the spins rotate. For $R$=Tb
and Dy these two vectors are perpendicular to each
other.\cite{Ken05} The coupling of a uniform electric polarization
\Pol\ to an inhomogeneous magnetization \boldmath${M}$\unboldmath\
is phenomenologically described by a term linear in the gradient
\boldmath$\nabla M$\unboldmath, the so-called Lifshitz invariant.
Such a coupling breaks the inversion symmetry of the crystal lattice in the
spiral magnetic state and induces the direction of the polarization
\boldmath$P\mbox{\unboldmath$=\gamma\chi M_{1}M_{2}[$}e
\mbox{\unboldmath$\times$} \kappa\mbox{\unboldmath$]$}$ \unboldmath,
where $\gamma$ and $\chi$ are a coupling constant and the dielectric
susceptibility, $M_{1}$ and $M_{2}$ are amplitudes of the
magnetic moments in directions perpendicular to \vecE.\cite{Mos06}
For $R=$Tb and Dy, \boldmath$\kappa$\unboldmath\ is parallel to the
$b$-axis and \vecE\ parallel to the $a$-axis, so that the
ferroelectric polarization induced below the spiral ordering
temperature $T_{C}$,  is parallel to the $c$-axis
(\ppc).\cite{Kim03a}

The application of magnetic field either parallel to the $a$- (\hpa)
or $b$-axis (\hpb) leads to a flop of the polarization from \ppc\ to
\ppa.  The flop in the polarization is interpreted as the flop of the
vector \vecE\ from the $a$- (spins within the $bc$-plane) to the
$c$-direction (spins within the $ab$-plane). For $R$=Tb these
field-induced flops of \Pol\ occur at critical field of \hca$\sim$8~T
and \hcb$\sim$4.5~T at 4~K for field parallel to the $a$- and $b$-axis
respectively and are associated with a first order transition from an
incommensurate (IC) low field magnetic ordering to a commensurate (CM)
high field phase. The $H$-$T$-phase diagram of \mbox{DyMnO$_{3}$}\ is
very similar to the one of \mbox{TbMnO$_{3}$}, with the same
characteristic flops of \Pol\ but at lower critical fields,
\hca$\sim$6.5~T and \hcb$\sim$1~T at 2~K.\cite{Kim05}

In this paper we show, that although the phase diagrams of $R$=Dy and Tb may
be qualitatively similar, their structural behavior at $H_{C}$ is completely
different. We find that for \mbox{DyMnO$_{3}$}\ the polarization flop is not
associated with a transition to a CM phase as in the case of $R=$Tb, but
rather the magnitude of the wave vector changes very little across $H_{C}$
for both \hpa\ and \hpb\ configurations. We argue that the magnitude of the
incommensurability for $R$=Dy does not lie sufficiently close to a CM value,
as opposed to $R$=Tb, making the IC high field phase energetically more
favorable.

For $R$=Dy, Mn-spins order to form a longitudinal spin density wave (SDW)
below $T_{N}$$\sim$39~K with wave vector \boldmath${\kappa}=%
\mbox{\unboldmath$\delta_{m}$}{b^{*}}$\unboldmath\ with \deltamn$\sim$
0.36...0.385\cite{Kim05} determined on the basis of lattice reflections with
\deltal=2\deltam\cite{Kim03b,Kim05,Fey06} that arise from a coupling of the
IC magnetic ordering to the lattice via a quadratic magneto-elastic coupling.
\cite{Wal80} With further cooling \deltamn\ decreases down to $T_C$=19~K,
where a second transition into the spiral phase occurs. Coincident with the
transition to a spiral phase a spontaneous electric polarization parallel to
the $c$-axis is found.\cite{Kim03a} Below \tndy=5~K Dy magnetic moments
order commensurately with propagation vector $\frac{1}{2}\mbox{
\boldmath$b^*$}$. \cite{Fey06}
In the study presented here, structural first and second harmonic
reflections related to the magnetic first harmonic reflections are
investigated.

Single crystalline samples were prepared using the floating zone
technique at the HMI.  Details of sample preparation and
characterization are given elsewhere.\cite{Fey06} Due to the high
neutron absorption cross section of Dy, in-field neutron diffraction
is not ideal to investigate field induced magneto-structural
transitions in \dymno. Rather to investigate the structural response
to the field induced polarization flop we have utilized in-field
synchrotron x-ray diffraction with the two field configurations \hpa\ and
\hpb.
Measurements with \hpa\ were performed at beamline X21 at the
National Synchrotron Light Source at Brookhaven National Laboratory
with a photon energy of 9.5~keV, using a 13~T Oxford cryomagnet with
vertical field.
Measurements in the \hpb\ configuration
were carried out at beamline BW5 at HASYLAB
with a photon energy of 100~keV using a 10~T Cryogenics cryomagnet with
horizontal field.
Magnetization measurements were performed with a 
Physical Properties Measurement System (PPMS) on a 14.5~mg
\dymno\ sample.
The spontaneous electric polarization as a function of field and
temperature was determined from the pyro-current recorded in a
PPMS-system 
using an
electrometer.

\begin{figure}[tbp]
\includegraphics[width=8cm]{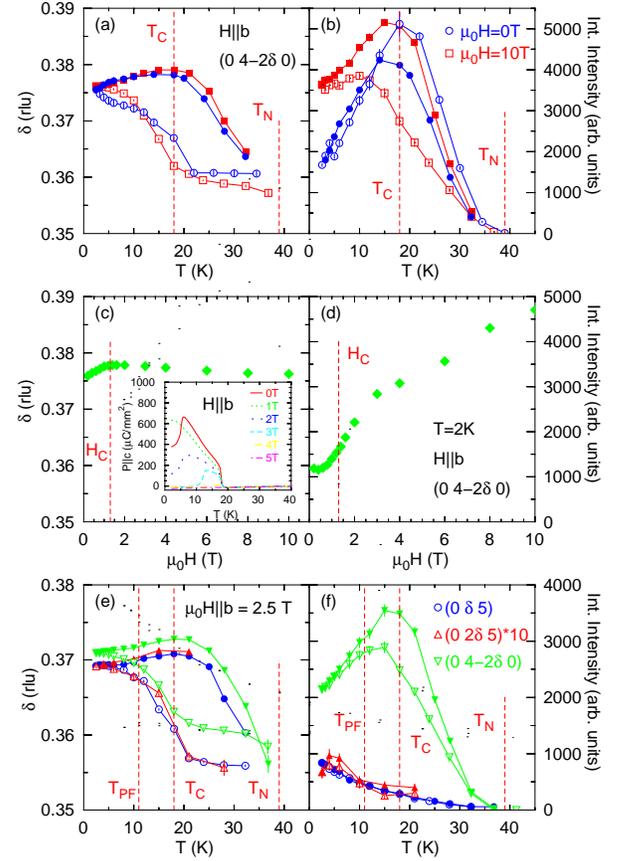}
\caption{(color online) Temperature dependence of (a) the incommensurability
$\protect\delta$ and (b) the respective intensities of the (0, 4-2$\protect
\delta$, 0) structural reflection. Data are shown for $\protect\mu_0 H = 0$
~T and $\mhpb = 10$~T for decreasing (open symbols) and increasing (closed
symbols) temperature. In panel (c) and (d), $\protect\delta$ and intensity
variation as function of magnetic field, respectively, are shown for a
sample temperature of $T=2$~K, in the same axes ranges as in (a) and (b). In
the inset, spontaneous electric polarization \ppc\ is shown as function of
temperature for magnetic field orientation \hpb. Temperature dependence of
(e) wave vector and (f) the respective intensities for $\mhpb = 2.5$~T and
wave vectors (0,\ $\protect\delta$,\ 5), (0,\ 2$\protect\delta$,\ 5) and
(0,\ 4-2$\protect\delta$,\ 0) for decreasing (open symbols) and increasing
(closed symbols) temperature. The (0,\ 2$\protect\delta$,\ 5) intensities in
(f) are multiplied by a factor of 10.}
\label{tdep_0t_10t}
\end{figure}

\begin{figure}[tbp]
\includegraphics[width=8cm]{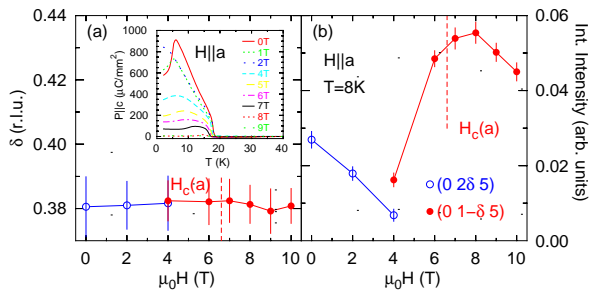}
\caption{(color online) Field dependences with field \hpa\ of (a) the
incommensurability and (b) the integrated intensity measured at $T=8$~K. The
error bars in (a) represent the HWHM of the superlattice reflection. In the
inset, spontaneous electric polarization \ppc\ is shown as function of
temperature for magnetic field \hpa.}
\label{tdep_025t_l}
\end{figure}

Measurements of the spontaneous electric polarization performed on the same
samples used in our diffraction experiments in magnetic field configurations
\hpa\ and \hpb\ (insets in Fig.~\ref{tdep_0t_10t}c and Fig.~\ref{tdep_025t_l}
a) confirm that spontaneous ferroelectric polarization is present below $
T_{C}$=19~K as already reported in Ref.~\onlinecite{Kim05}. The application
of \hpb\ below \tc\ results in the suppression of \ppc\ 
(inset of Fig.~\ref{tdep_0t_10t}c). 
Kimura \etal show in addition that this decrease is
accompanied by an increase in \ppa\ indicative of the flop in the
ferroelectric polarization.\cite{Kim05}


%

In Fig.~\ref{tdep_0t_10t}a-b we show the temperature dependence of the wave
vector and integrated intensity of the second harmonic reflection (0, 4-2$
\delta$, 0), both for decreasing and increasing temperature. The dependences
are compared for zero field (\ppc), and $\mhpb=10$~T (\ppa). In both data
sets, we find a significant hysteresis in the intensity as well as in the
magnitude of $\delta$ around $T_C$=18~K, which is associated with the onset
of ferroelectricity. However, despite the fact that \Pol\ lies along
different axes for 0 and 10~T, we observe no significant change between
these two measurements. In Fig.~\ref{tdep_0t_10t}c-d, we show the field
dependence of $\delta$ and of the integrated intensity of the same
reflection at $T$=2~K. Here we find a small initial increase of $\delta$ up
to \mhcb=1~T where the flop in polarization is found while the intensity of
the same reflection shows a steady increase with increasing field. Above \hcb
, we find a small decrease of $\delta$ with increasing field. This behavior
is in sharp contrast to \mbox{TbMnO$_{3}$}\ where $\delta$ varies slowly
with increasing field and locks in above \hcb\ into a CM value of \boldmath$
\kappa\mbox{\unboldmath$=\frac{1}{4}$}\mathbf{b^{*}}$\unboldmath\ at \hcb.
\cite{Ali06a}

In Fig.~\ref{tdep_0t_10t}e-f, the temperature dependence of $\delta$ and the
intensities of first and second harmonic reflections is shown for data
measured in field cooling with $\mhpb=2.5$~T. Here we find a different
behavior in the intensities of IC reflections. For wave vector along the
\bdirection\, (0, 4-2$\delta$, 0), a strong hysteresis is observed in its
intensity as a function of temperature, whereas this hysteresis is absent
for wave vectors that are mainly along the \cdirection\ ( (0, $\delta$, 5)
and (0, 2$\delta$, 5)) (Fig.~\ref{tdep_0t_10t}f). Nevertheless, $\delta$ and
its hysteresis are the same for all reflections (Fig.~\ref{tdep_0t_10t}e).
The described intensity behavior is similar to what we have recently
observed in zero field temperature dependent measurements using resonant
x-ray scattering from a single crystal of DyMnO$_{3}$.\cite{Pro07} 
In these
measurements a similar hysteresis was observed to be associated with the
induced ordering of Dy-spins with the same propagation vector as that for
the Mn spin ordering.
Finally for
this field configuration we note that at 2.5~T, the polarization flop from
\ppa\ to \ppc\ is expected with increasing temperature at $T_{PF} \sim 12$~K
(inset in Fig.~\ref{tdep_0t_10t}c). However our diffraction measurements
find no anomaly either on the temperature dependence of the wave vector or
in the intensities at this temperature as it was found in \mbox{TbMnO$_{3}$}.

We now turn our attention to measurements conducted in the \hpa\
configuration. In Fig.~\ref{tdep_025t_l}a-b the field dependence of the
intensity and wave vector of the first and second harmonic reflections 
(0, 1-$\delta$, 5) and (0, 2$
\delta$, 5) is shown, measured at $T=8$~K above
the ordering temperature of Dy of $\tndy\sim 5$~K. The second harmonic
reflection decreases in intensity with increasing field and vanishes at 5~T,
while the first harmonic appears only at a field of 4~T and saturates in
intensity at 7~T. This behavior is not directly related to the polarization flop but reflects a variation of the modulation direction of the strain wave leading to the superlattice reflections. 
However the most remarkable behavior is found for the
field dependence of the wavevector. Here, the wave vector of the first
harmonic reflection does not change significantly through the polarization
flop transition at \mhca$\sim 6.5$~T.\cite{Kim05} Again this is in sharp
contrast to the behavior of \tbmno, where a discontinuous
transition to a CM phase is found.

\begin{figure}[tbp]
\includegraphics[width=8cm]{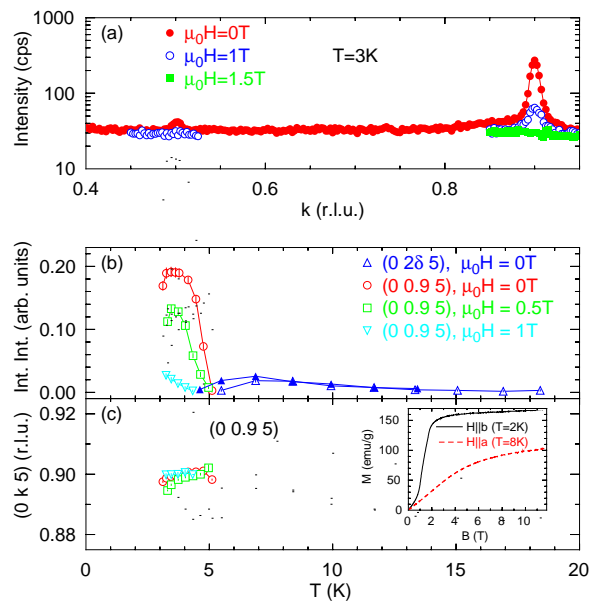}
\caption{(color online) (a) Scans along (0, $k$, 5) at $T=3$~K for
different applied magnetic fields with intensity in logarithmic
scale. (b) Temperature dependence as function of field \hpa\ of the
integrated intensity of the superlattice reflections (0, 0.9, 5) in
the Dy ordered phase and the second harmonic reflection 
$(0,\ 2\protect\delta,\ 5)$ due to Mn order. Open and closed symbols
represent increasing and decreasing temperature, respectively. In (c)
the wave vector of the (0, 0.9, 5) reflection is shown as function of
temperature for different fields. The inset shows magnetization data of
\mbox{DyMnO$_{3}$}\ for \hpa\ and \hpb.}
\label{tdep_0t_tt}
\end{figure}

Below 6K, Dy spins order commensurately with $\delta_{m}^{Dy}=\frac{1}{2}$.
Previously we have argued that the lattice distortion associated with this
magnetic ordering is not CM but rather IC with $\delta_{l}\sim0.1\mathbf{
b^{*}}$.\cite{Fey06} In our measurements we found the half-integer
magnetic reflection to be extremely weak and we focused our attention to the
lattice $\delta_{l}\sim0.1$ satellite measured at $(0,\ 0.9,\ 5)$. In Fig.~
\ref{tdep_0t_tt}a we show a series of scans at different fields applied
along the $a$-axis at 3~K which show the rapid suppression of the $
\delta_{l}\sim0.1$ satellite which vanishes for fields $\mhpa
>1$~T. The CM magnetic reflection $(0,\ 0.5,\ 5)$ is only measurable at zero
field and can not longer be observed at 1~T. The temperature dependence of
the intensity and wavevector of the $(0,\ 0.9,\ 5)$ reflection is shown in
Fig.~\ref{tdep_0t_tt}b-c, revealing the rapid suppression of the intensity
of this reflection at low temperature in accordance with the appearance of
ferromagnetic order (inset Fig.~\ref{tdep_0t_tt}c), indicating an easy axis
along the $b$-direction. The suppression of this reflection with magnetic
field is analogous to a similar behavior found for $R=$Tb which coincided
with the ferromagnetic ordering of Tb-spins for the same field size and
configuration.\cite{Ali06a}

The polarization flop is driven by the field-induced flop of the
axis of rotation of the magnetic spiral \vecE.  Its direction is
determined by magnetic anisotropy terms, which to lowest order of
the free energy expansion in powers of magnetization of Mn spins
have the form, $\sum_{\alpha =a,b,c}a_{\alpha } M_{\alpha }^{2}$.
Phenomenologically the spin flop transition results from the field
dependence of the coefficients $a_{\alpha}$. Below the critical
field  $H_{C}$, $a_{b}<a_{c}<a_{a}$, which favors spins rotating in
the $bc$ plane ($e\Vert a$), while above $H_{C}$,
$a_{b}<a_{a}<a_{c}$, favoring the rotation in the $ab$-plane
($e\Vert  c$).

From the perspective of symmetry there is no restriction that the
high field spiral phase must be CM. In this view the fact that there
is a magneto-elastic phase transition to a CM phase associated with
the flops in the polarization for $R=$Tb would appear to be a
special case, especially when compared to $R=$Dy where we find no
such transition at $H_{C}$. The difference in behavior between these
two multiferroics is not of fundamental nature but rather simply
lies in the magnitude of the incommensurability. For $R$=Tb,
$\delta_{m}$=0.28~r.l.u. is close to the CM value of $\frac{1}{4}$.
For this CM value of the wave vector the amplitude of the Mn
magnetic moment is not modulated\cite{Ali06a} and may thus be
energetically more favorable than an IC amplitude modulated phase.
In the phenomenological approach the CM state with $\delta$ =
0.25~r.l.u. is stabilized by $M^4$ terms in the Landau expansion.
For $R$=Dy the value of $\delta$=0.38~r.l.u. is further away from a
CM value ($\frac{1}{3}$) and thus a transition to it is not
favorable energetically. Therefore the Mn spin spiral may indeed
flop as expected but without a change in $\delta$.
\footnote{%
We note that a value of $\delta=\frac{1}{3}$ is observed for Tb$_{1-x}$Dy$%
_{x}$MnO$_{3}$, see Arima \etal\ Phys. Rev. Lett. \textbf{96},
097202 (2006).} Clearly here the modulation period in real space is
much shorter, and the amplitude of the Mn magnetic moment is
modulated.

A significant difference between the two multiferroics is found in the
behavior of the complex magnetic ordering of $R$-ions. For $R$=Tb, Tb spins
are induced to order along the $a$-axis below $T_{N}$ with the same
periodicity as Mn-spins and below \tntb=7~K they order separately with $
\delta_{m}^{Tb}=$0.42~r.l.u. At low temperatures, when \hpa$\sim$1~T,
Tb-spins are aligned ferromagnetically, while for field along the $b$-axis, $
\delta_{m}^{Tb}$ jumps discontinuously to a CM value of $\frac{1}{3}$ at
1.75~T.\cite{Ali06a} The behavior for $R$=Dy is much simpler, as below \tndy
=6~K the Dy spins order with $\delta_{m}^{Dy}=\frac{1}{2}$.
Here, the suppression of the IC-reflections together with the magnetization
data indicates a melting of the antiferromagnetic Dy ordering, a behavior
different to \mbox{TbMnO$_{3}$}.

On the basis of the CM ordering of Mn-spins for the \ppa\ phase for $R$=Tb
it has been suggested that ferroelectricity may arise in the absence of a
spiral magnetic ordering. Here an exchange striction mechanism proposed from
competing ferromagnetic and antiferromagnetic super-exchange interactions
predicts a \ppa\ phase for a CM ordering with $\delta=\frac{1}{4}$. This
model holds strictly for a CM ordering and suggests that for $R$=Dy the
spiral phase must be maintained at high fields to support a ferroelectric
state.

The smooth magnetic field dependence of the spiral wave vector in
\mbox{DyMnO$_{3}$}\ at the spin flop transition may explain the
large increase of the dielectric constant $\varepsilon _{a}$,
observed in this material.\cite{Got04,Kim05} If higher-order terms
in the Landau expansion of free energy could be neglected, then at
the spin flop transition the magnetic excitation spectrum of the
spiral would acquire a zero mode, since for $a_{a}\left(
H_{C}\right) =a_{c}\left(H_{C}\right)$ there is a freedom to rotate
the spiral plane around the $b$ axis. This mode can be excited by
electric field $E\Vert a$ normal to the spiral $bc$-plane and is the
electromagnon studied in Ref.~\onlinecite{Pim06}. Its softening at
$H=H_{C}$ would result in divergence of static dielectric
susceptibility $\varepsilon _{a}$.  In reality, due to higher-order
terms  in the Landau expansion the spin flop transition is of first
order, the softening of the magnetic mode is not complete and the
peak value of the dielectric constant is finite. Still, in
\mbox{DyMnO$_{3}$}\ this transition is close to a second-order one
in the sense that the spirals above and below the critical field are
essentially the same except for their orientation. The softness of
the spiral magnetic ordering at the critical field may be the reason
behind the large magnetocapacitance observed in \mbox{DyMnO$_{3}$}\,
which becomes truly gigantic close to the tricritical point at the
crossing of the collinear and two spiral phases with \ppc\ and \ppa,
where higher-order terms are small. In \mbox{TbMnO$_{3}$}\ the spin
flop transition is strongly discontinuous due to the concomitant
IC-CM transition, which limits the growth of the dielectric constant.

In summary, in-field synchrotron X-ray diffraction measurements from a
\mbox{DyMnO$_{3}$}\ single crystal have shown that there is no change of the
wave vector $\delta$ associated with the flop of the ferroelectric
polarization $\mathbf{P}$ at $H_{C}$ for both \hpa\ and \hpc. This is in
sharp contrast to similar measurements reported for \mbox{TbMnO$_{3}$}\ were
a transition to a CM phase is found at $H_{C}$ for the same field
configurations. We argue that the magnitude of the incommensurability for $R$
=Dy does not lie sufficiently close to a CM value, as opposed to $R$=Tb,
making the IC high field phase energetically more favorable.


\begin{acknowledgments}
We would like to thank C.S. Nelson for the assistance at the experiment at
NSLS. Work at Brookhaven was supported by the U.S. Department of Energy,
Division of Materials Science, under Contract No. DE-AC02-98CH10886.
\end{acknowledgments}


\end{document}